\documentclass[10pt,conference]{IEEEtran}
\usepackage{amsmath, amssymb, graphicx, cite,comment}
\usepackage{balance}

\usepackage{epstopdf, soul, color}

\newtheorem{thm}{Theorem}

\newtheorem{defin}{Definition}

\DeclareMathOperator*{\argmax}{arg\,max}
\DeclareMathOperator*{\argmin}{arg\,min}

\newcommand{\cX}{{\cal{X}}}
\newcommand{\cY}{{\cal{Y}}}
\newcommand{\cW}{{\cal{W}}}
\newcommand{\cF}{{\cal{F}}}

\IEEEoverridecommandlockouts 

\title{Multicasting Energy and Information Simultaneously}
\author{%
   \IEEEauthorblockN{Ting-Yi~Wu\IEEEauthorrefmark{1},
                     Anshoo Tandon\IEEEauthorrefmark{2},
                     Lav~R.~Varshney\IEEEauthorrefmark{3},
                     and Mehul Motani\IEEEauthorrefmark{2}}
   \IEEEauthorblockA{\IEEEauthorrefmark{1}%
                     Sun Yat-Sen University, 
                     wutingyi@mail.sysu.edu.cn}
   \IEEEauthorblockA{\IEEEauthorrefmark{2}%
                     National University of Singapore,  
                     \{anshoo.tandon@gmail.com, motani@nus.edu.sg\}}
    \IEEEauthorblockA{\IEEEauthorrefmark{3}%
                     University of Illinois at Urbana-Champaign, 
                     varshney@illinois.edu}
}

\begin{document}
\maketitle

\begin{abstract}
Communication systems for multicasting information and energy simultaneously to more than one user are investigated. 
In the system under study, a transmitter sends the same message and signal to multiple receivers over distinct and independent channels.  The fundamental communication limit under a received energy constraint, called the multicast capacity-energy function, is studied and a single-letter expression is derived.  This is based on coding theorems for compound channels. 
The problem of receiver segmentation, where receivers are divided into related groups, is also considered.
\end{abstract}

\section{Introduction}
Over the last decade, energy harvesting communication systems have not only been an attractive research topic, but have also seen considerable progress towards practical design and implementation. 
Wireless power delivery, rather than supplying power via wire or battery, may
save significant cost, especially where battery replacement is difficult \cite{TongLWZ2010,LeeLL2012}.  

The idea of \emph{simultaneous} information/energy transmission (SIET) to an energy harvesting receiver provides the possibility of trading between information transfer and energy delivery \cite{ZhangMH2015,ClerckxZSNKP2018}. 
There is a fundamental limit for point-to-point communication, when the transmitter must also transfer some minimal amount of energy to a receiver, as characterized by a \emph{capacity-energy} function \cite{Varshney2008} e.g.\ for discrete memoryless channels (DMCs) and Gaussian channels.  This capacity-energy function can also be computed for SIET over noisy coupled-inductor circuits, modeled as a frequency-selective channel with additive white Gaussian noise \cite{GroverS2010}. Suboptimal strategies such as time-switching, where the transmitter switches between information delivery and energy delivery phases, can also be characterized \cite{Varshney2012,LiuZC2013}. In \cite{TandonMV2016}, the constrained capacity over DMCs was derived when each \emph{subblock} in a codeword must carry sufficient energy to avoid receiver energy outage. These subblock energy-constrained codes were generalized to \emph{skip-sliding window codes} in \cite{WuTVM2017}.

\begin{figure}[t]
  \centering
    \includegraphics[width=3.4in]{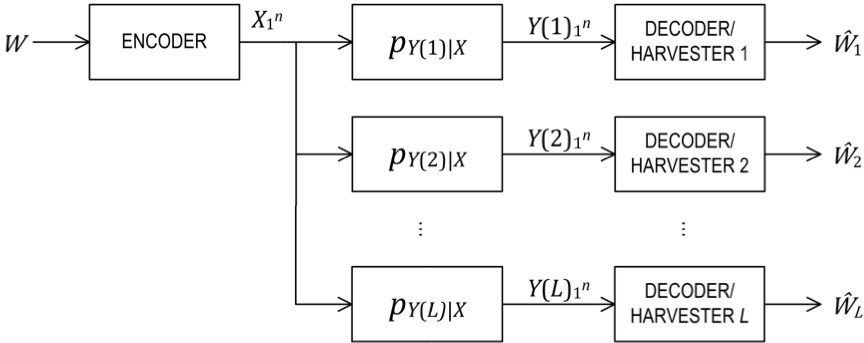}\vspace{-0.1in}
  \caption{The multicast setting: communicating the same signal over distinct noisy channels to several energy-harvesting receivers.}
  \vspace{-0.15in}
  \label{fig:multicast}
\end{figure}

There have been numerous multiterminal extensions to the basic energy/information transmission problem \cite{FouladgarS2012,ZhangH2013}.
In \cite{FouladgarS2012}, SIET over multiple access channels was analyzed, whereas 
\cite{ZhangH2013} studied broadcasting information/energy messages from a base station to energy-harvesting nodes and information-retrieving nodes.  
Previous work has, however, not studied the practically important \emph{multicast} setting of Fig.~\ref{fig:multicast}, where the same message is transmitted over distinct and independent channels.  

Although the idea of multicasting PoWiFi has become prominent in the communication systems literature \cite{TallaKRNSG2017} (leading to much public attention 
and even a start-up company), these ideas have not previously been tied to fundamental principles concerning physical layer capacity.  Note that beacon signals form a significant bulk of wireless protocols, and so the multicast problem (as compared to the broadcast problem) is often present.

Tree-structured multicast is simply a compound channel, since requiring reliable 
communication over several channels is equivalent to requiring reliable communication over an unknown channel.  
Recall the capacity of a compound DMC does not increase if the decoder (but not the encoder) knows which channel is chosen from a given family of channels \cite{LapidothN1998}.  In defining compound capacity 
here, in addition to requiring a small error probability over each channel, we also require a certain minimum energy delivery.

\begin{figure}[t]
  \centering
    \includegraphics[width=3.5in]{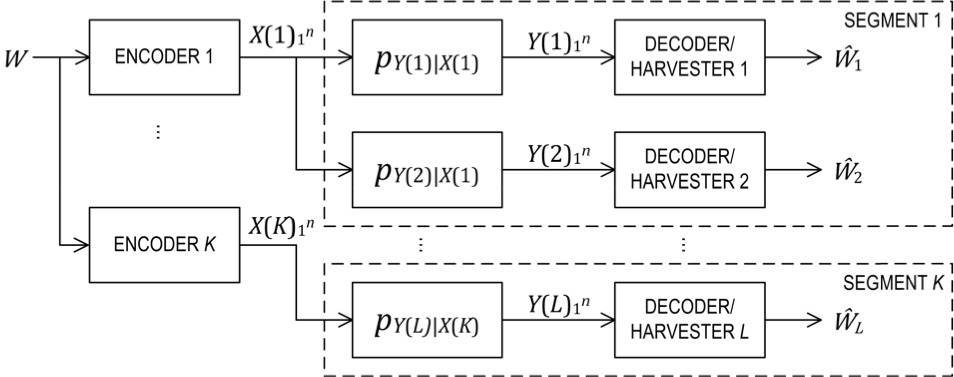}\vspace{-0.1in}
  \caption{Communicating the same message and a few signals over distinct noisy channels to grouped energy harvesting receivers, the segmentation problem.}
  \vspace{-0.15in}
  \label{fig:multicast_group}
\end{figure}

In SIET communication systems, one may consider partitioning receivers into groups, based on their distance from the base-station\cite{ZengCZ2017} or energy requirements.
Therefore, beyond determining the compound capacity-power function, we also consider \emph{receiver segmentation},
where we group channels into different categories, so the same signaling scheme is used for members of the same category, as in Fig.~\ref{fig:multicast_group}.  
This is equivalent to optimally quantizing the parameter space of the compound channel under given
information/energy requirements.  

In summary, this work defines energy/information multicast communication and derives a single-letter expression for its capacity under a received energy constraint. In addition, a brief discussion of the segmentation problem is presented.  

\section{Point-to-Point Capacity-Energy Function}
Consider a DMC with transition probability $p_{Y|X}(x|y)$, in which 
$x$ and $y$ are drawn from the input alphabet $\cX$ and the output alphabet $\cY$ with distribution $p_{X}(x)$ and $p_{Y}(y)$, respectively. 
Let $b(y)$ be an energy function that maps a channel output $y\in\cY$ into a nonnegative real value. 
When channel inputs described by random variables $X_1^n=(X_1, X_2, \ldots, X_n)$ are transmitted through the channel, 
the corresponding channel outputs described by random variables $Y_1^n=(Y_1, Y_2, \ldots, Y_n)$ 
contain the average energy  
\begin{equation}
E[b(Y_1^n)]=\sum_{y_1^n = (y_1,y_2,\ldots,y_n)\in \cY^n}b(y_1^n)p_{Y_1^n}(y_1^n),
\end{equation}
where $b(y_1^n)\triangleq\sum_{i=1}^n b(y_i)$. 

Consider the point-to-point case: 
an $n$-letter transmission which guarantees at least expected energy $nB$ being received 
can be evaluated by the $n$th capacity-energy function as
\begin{equation}
C_n(B)=\max_{X_1^n:E[b(Y_1^n)]\ge nB}I(X_1^n;Y_1^n).
\end{equation}
As mentioned in \cite{Varshney2008}, 
the capacity of the channel under constraint $(b, B)$, denoted $C(B)$, 
can be single-letterized as 
\begin{equation}
C(B)=C_1(B). 
\end{equation}

\section{Multicasting Capacity-Energy Function}
Let us consider the multicast SIET setup in Fig.~\ref{fig:multicast}.
The message alphabet $\cW$ is encoded by the function $f$, which maps each message word $w\in \cW$ 
into a codeword $x_1^n=f(w)\in\cX^n$. 
The source broadcasts the generated codeword to $L$ distinct receivers over independent DMCs and the $\ell$th client receives $y(\ell)_1^n\in\cY(\ell)^n$ 
based on the transition probability $p_{Y(\ell )_1^n|X_1^n}(y(\ell)_1^n|x_1^n)$. 
Since the channel is memoryless, $p_{Y(\ell )_1^n|X_1^n}(y(\ell)_1^n|x_1^n)=\prod_{i=1}^n p_{Y(\ell )|X}(y(\ell)_i|x_i)$. 
The $\ell$th client decodes $y(\ell)_1^n$ into $\hat{W}_{\ell}\in\cW$ by its own decoder $g_\ell(\cdot )$ and 
an error occurs when $\hat{W}_\ell\neq W$. 
Let $\Theta=\{1,2,\ldots,L\}$ be the set of $L$ receivers. 
The notion of error probability follows. 
\begin{defin}
The error probability for the multicast channel, denoted by $\eta$, is defined 
as the worst average error probability over all channels:
\begin{equation}
\eta=\max_{\ell\in\Theta}\Pr[W\neq\hat{W}_\ell].
\end{equation} 
\end{defin}

The $\ell$th client receives energy $E[b_{\ell}(Y(\ell)_1^n)]$ on average, 
where the energy harvesting function $b_{\ell}(\cdot)$ may be different for different receivers. 
The energy constraint vector $\vec{B}=(B_1,B_2,\ldots,B_L)$ is an $L$-tuple vector, 
which defines the least energy requirement for each channel, i.e.,
\begin{equation}\label{eqn:energy-constraint}
E[b_{\ell}(Y(\ell)_1^n)]\ge nB_\ell, \, \forall \ell\in\Theta. 
\end{equation}
An input that satisfies \eqref{eqn:energy-constraint} is said to be $\vec{B}$-admissible. 
Now we define the operational compound capacity in terms of an energy constraint vector $\vec{B}$. 
\begin{defin}
An encoder $f$ maps a message $w\in\cW$ into a block of length $n$ with a rate $R=\frac{1}{n}\log |\cW|$. 
Given $0\leq \epsilon < 1$, 
a non-negative rate $R$ is said to be an $\epsilon$-achievable rate for multicast channels $\{p_{Y(\ell )|X}\}_{\ell\in\Theta}$
with an energy constraint vector $\vec{B}$ if
there exists a block code with error probability $\eta<\epsilon$ such that \eqref{eqn:energy-constraint} holds 
and its rate exceeds $R-\delta$ for all $\delta>0$ when $n$ is sufficiently large.
$R$ is an achievable rate if it is $\epsilon$-achievable for all $\epsilon>0$. 
The supremum of the achievable rate is called the operational capacity of the multicast channel, denoted $C_{\mathrm{O}}(\vec{B})$.
\end{defin}

To evaluate the capacity, let us consider informational quantities. 
Let all feasible channel input distributions of dimension $n$ with the $\ell$th energy constraint be $\cF_n(\ell)$, i.e.,
\begin{equation}
\cF_n(\ell)=\left\{F_{X^n_1}: E[b_{\ell}(Y(\ell)_1^n)]\ge nB_\ell \right\},
\end{equation} 
then the set all $\vec{B}$-admissible inputs is $\cF_n=\bigcap_{\ell\in\Theta}\cF_n(\ell)$.
The multicast capacity-energy function is then defined as follows.
\begin{defin}
Given a positive integer $n$ and an energy constraint vector $\vec{B}$, the $n$th multicast capacity-energy function $C_n(\vec{B})$ is defined as 
\begin{equation}\label{eqn:nth-cefunction}
C_n(\vec{B})=\max_{X_1^n\in\cF_n}\min_{\ell\in\Theta}I(X_1^n;Y(\ell)_1^n),
\end{equation}
where $I(\cdot;\cdot)$ is the mutual information. 
The domain ${\cal B}$ of $C_n(\vec{B})$ is all feasible energy constraints:
\begin{multline}\label{def:domain}
{\cal B}\triangleq \Bigg\{(B_1,\ldots ,B_L):  B_\ell=\vec{q} \cdot [P(\ell)] \cdot \vec{b}_\ell \quad \forall\ell \in \Theta\\
\mbox{ for all nonnegative }\vec{q}=\left[q_1\ \cdots\ q_{|\cX|}\right] \mbox{ s.t. } \sum_{i=1}^{|\cX|}q_i=1\Bigg\},
\end{multline} 
where $\vec{b}_\ell$ is a column vector of the $[b_{\ell}(y)]_{y\in\cY}$ and $[P(\ell)]$ is the transition matrix $[p_{Y(\ell)|X}(y|x)]_{(x,y)\in\cX\times\cY}$.
The supremum rate of \eqref{eqn:nth-cefunction} defines the multicast capacity-energy function as 
\begin{equation}\label{eqn:cefunction}
C(\vec{B})=\sup_n \frac{1}{n}C_n(\vec{B}).
\end{equation}
\end{defin}
Now we have a coding theorem. 
\begin{thm}\label{thm:capacityfunction}
$C_{\mathrm{O}}(\vec{B})=C(\vec{B})$.
\end{thm}
\begin{IEEEproof}
Multicast energy/information communication can be treated as a compound channel with a constrained set of possible input distributions $\cF_n$. 
Hence, the achievability and converse can be derived from random coding arguments and Fano's inequality, respectively. 
See \cite[Thm.\ 7.1 \& Rem.\ 7.3]{ElGamalK2011}.
\end{IEEEproof}

Just as with traditional compound capacity, the compound capacity-energy function is not generally equal to the minimum of the individual 
capacity-energy functions of the channels constituting the set since different channels may have different optimal input distributions.  
That minimum is, however, an upper bound on the compound capacity-energy function.
Exchanging minimum and maximum operations is possible when the optimal input 
distributions in the set are identical.  

Although the multicast capacity-energy function provides an explicit form for capacity, 
finding the supremum of $n$-dimensional mutual information is difficult. 
Here we aim to provide a single-letter expression for the capacity. 

For notational convenience, we consider the setting where the energy harvesting functions $b_{\ell}(\cdot)$
are all identically $b(\cdot)$ and all receivers have the same energy constraint $B$:
$B=B_1=B_2=\cdots=B_L\leq B_{\max}$,
where $B_{\max}$ is given as
\begin{equation}
B_{\max}=\max_{\substack{\vec{q}=\left[q_1\ \cdots\ q_{|\cX|}\right] \\\mbox{ s.t. } \sum_{i=1}^{|\cX|}q_i=1\, \&\, q_i\ge 0\,\forall i}}\min_{\ell}\vec{q}\cdot [P(\ell)]\cdot\vec{b}.
\end{equation}
Then $C_n(B)$ is a function of a scalar rather than a vector.  Extension to the general case follows
directly.

It is trivial that $C_n(B)$ is non-increasing along $B$.  
The following theorem also shows that $C_n(B)$ is concave. 
\begin{thm}\label{thm:concavity}
$C_n(B)$ is a concave function of $B$ for $0\le B\le B_{\max}$.
\end{thm}
\begin{IEEEproof}
Let $(X_i)_1^n$ and $\{Y_i(\ell)_1^n\}_{\ell\in\Theta}$ be the channel input of length $n$ and the corresponding channel outputs 
which achieves $C_n(B_i)$ for $i=\{1,2\}$.
By definition, $E[b(Y_i(\ell)_1^n)]\ge nB_i$ for every $i$ and $\ell$.
Let $X_1^n$ be the channel input with the distribution $\alpha p_{(X_1)_1^n}+\beta p_{(X_2)_1^n}$ and $\{Y(\ell)_1^n\}_{\ell\in\Theta}$ 
be the corresponding channel outputs for some $\alpha\in[0,1]$ and $\beta=1-\alpha$.
The energy received from $Y(\ell)_1^n$ for the $\ell$th channel can be lower-bounded as
\begin{IEEEeqnarray}{rCl}
E[b(Y(\ell)_1^n)]&=&\alpha E[b(Y_1(\ell)_1^n)]+\beta E[b(Y_1(\ell)_1^n)]\\
&\ge& \alpha nB_1+\beta nB_2=n(\alpha B_1+\beta B_2).
\end{IEEEeqnarray}
Therefore, the channel input $X_1^n$ is $(\alpha B_1+\beta B_2)$-admissible.
By definition, we have the following inequalities for all $\ell\in\Theta$.
\begin{equation}
C_n(\alpha B_1+\beta B_2)\ge\min_{\ell\in\Theta}I(X_1^n;Y(\ell)_1^n).\label{eqn:concave-1}
\end{equation}
Since the mutual information is a concave function of the input distribution, 
\begin{multline}\label{eqn:concave-2}
I(X_1^n;Y(\ell)_1^n)\ge\\
\alpha I((X_1)_1^n;Y_1(\ell)_1^n)+\beta I((X_2)_1^n;Y_2(\ell)_1^n)
\end{multline}
for all $\ell\in\Theta$.
Combining \eqref{eqn:concave-1} and \eqref{eqn:concave-2} validates the concavity: 
\begin{align}
&C_n(\alpha B_1+\beta  B_2)\\
&\ge \min_{\ell\in\Theta}\left\{\alpha I((X_1)_1^n;Y_1(\ell)_1^n)+\beta I((X_2)_1^n;Y_2(\ell)_1^n)\right\}\\
&\ge \alpha \min_{\ell\in\Theta}\left\{I((X_1)_1^n;Y_1(\ell)_1^n)\right\}+\beta \min_{\ell\in\Theta}\left\{ I((X_2)_1^n;Y_2(\ell)_1^n)\right\}\\
&= \alpha C_n(B_1)+\beta C_n(B_2).
\end{align}
\end{IEEEproof}

The following theorem shows that the $n$th multicast capacity-energy function can be evaluated as a single-letter expression. 
\begin{thm}\label{thm:letterization}
For any DMC, $C_n(B)=nC_1(B)$ for all $n\in\mathbb{Z}^+$ and $0\le B\le B_{\max}$.
\end{thm}
\begin{IEEEproof}
To prove the equality, we start with the case of ``$\le$''.
Let $X_1^n$ be the channel input with the corresponding outputs $\{Y(\ell)_1^n\}_{\ell\in\Theta}$ which achieves $C_n(B)$, 
then we have $E[b(Y(\ell)_1^n)]\ge nB$ for all $\ell\in\Theta$. 
We also let $X^*$ be the channel input with the corresponding outputs $\{Y^*(\ell)\}_{\ell\in\Theta}$ such that 
$p_{X^*}=\sum_{i=1}^n \frac{1}{n}p_{X_i}$.
Due to the memoryless, We have 
\begin{IEEEeqnarray}{rCl}
E[b(Y^*(\ell))]&=&\frac{1}{n}\sum_{i=1}^nE[b(Y(\ell)_i)]\\
&=&\frac{1}{n}E[b(Y(\ell)_1^n)]\ge B
\end{IEEEeqnarray}
and $I(X_1^n;Y(\ell)_1^n)\le\sum_{i=1}^nI(X_i;Y(\ell)_i)$ for all $\ell\in\Theta$.
Hence
\begin{IEEEeqnarray}{rCl}
C_n(B)&=&\min_{\ell\in\Theta}I(X_1^n;Y(\ell)_1^n)\\
&\le&\min_\ell \sum_{i=1}^nI(X_i;Y(\ell)_i)\\
&=&n\min_\ell \frac{\sum_{i=1}^nI(X_i;Y(\ell)_i)}{n}\\
&\le&n\min_\ell I(X^*;Y^*(\ell))\label{eqn:singleletter-1}\\
&\le&nC_1(B),\label{eqn:singleletter-2}
\end{IEEEeqnarray}
where \eqref{eqn:singleletter-1} follows from \eqref{eqn:concave-2} and \eqref{eqn:singleletter-2} follows from the definition. 

For the case of ``$\ge$'',
we let $X$ be the channel input with the distribution $p_X$ and the corresponding outputs $\{Y(\ell)\}_{\ell\in\Theta}$ achieving $C_1(B)$.
A sequence of channel inputs $X_1^n=(X_1,X_2,\cdots,X_n)$ is created, in which each $X_i$ is i.i.d. to $X$. 
Let $\{Y(\ell)_i^n\}_{\ell\in\Theta}$ be the corresponding channel outputs of $X_i^n$.
By definition, $E[b(Y(\ell)_1^n)]=\sum_{i=1}^nE[b(Y(\ell))]\ge nB$.
Therefore
\begin{IEEEeqnarray}{rCl}
C_n(B)&\ge&\min_{\ell\in\Theta}I(X_1^n;Y(\ell)_1^n)\\
&=&\min_{\ell\in\Theta}\sum_{i=1}^nI(X;Y(\ell))\\
&=&nC_1(B).
\end{IEEEeqnarray}
\end{IEEEproof}

Combining Theorems \ref{thm:capacityfunction} and \ref{thm:letterization} shows 
the capacity of the multicast SIET as in Fig.~\ref{fig:multicast} is single-letterized as
\begin{equation}\label{eqn:capacity}
C_{\mathrm{O}}(B)=C_1(B).
\end{equation}

Although single-letterization was shown for identical energy harvesting functions and constraints, 
the derivations are still valid for unequal constraints\footnote{
The single-letterization for unequal constraints $\vec{B}$ and different energy functions $b_\ell$ 
is given in Appendix \ref{append:letterization_general}}. 

\subsection{Numerical Example}
The multicast capacity-energy function is evaluated for an example.
Considering a case of $L=2$, in which both channels are binary, denoted by $\Theta=\{\mathrm{BSC},\mathrm{Z}\}$. 
The first channel is a binary symmetric channel (BSC) with crossover probability $\epsilon$; 
the second channel is a Z-channel with a $1$ to $0$ probability $\epsilon_0$. 
Let the energy function $b(\cdot)$ be identically defined as the Hamming weight of the output 
and have common energy delivery requirement $B$. 

\begin{figure}
  \centering
    \includegraphics[width=3.5in]{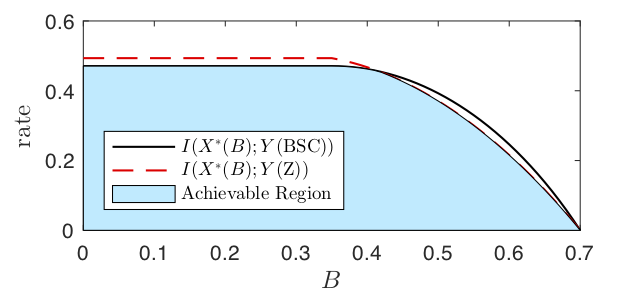}\vspace{-0.1in}
  \caption{An example of two-receivers multicast communication. Two binary channels were considered: a BSC with crossover probability $0.12$ and
  a Z-channel with a $1$ to $0$ probability $0.3$.}
  \vspace{-0.15in}
  \label{fig:ex}
\end{figure}

Let $X^*(B)$ be the common capacity-achieving input distribution, whose computation is detailed elsewhere \cite{WuTVM2018_arXiv}.  For given values of $\epsilon$ and $\epsilon_0$, Fig.~\ref{fig:ex} individually plots
the mutual informations of both channels with input $X^*(B)$. 
Since the multicast capacity is dominated by the worst channel, $C_{\mathrm{O}}(B)$ is simply the minimum of the curves.

\section{Gaussian Channels}
Now consider multicast over $L$ discrete-time, continuous-alphabet, memoryless channels.
Given an input $x$, the received energy at receiver $\ell$ is
\begin{equation}
\rho_\ell(x)=\int Q_\ell(y|x)b_{\ell}(y)dy,
\end{equation}
where $Q_\ell(y|x)$ is the transition probability of the $\ell$th channel. 
Let $F(x)$ be the cumulative distribution function (cdf) of $X$; 
the expectation of $\rho_\ell(X)$ is:
\begin{IEEEeqnarray}{rCl}
E[\rho_{\ell}(X)]&=&\int\rho_\ell (x)dF(x)\\
&=&\int\int Q_\ell(y|x)b_{\ell}(y)dydF(x)\\
&=&\int\left(\int Q_\ell(y|x)dF(x)\right)b_{\ell}(y)dy\\
&=&E[b_{\ell}(Y(\ell))],
\end{IEEEeqnarray}
which indicates that the output energy constraint can be considered as the function of $\rho_\ell(x)$ at the input side.
Theorem \ref{thm:letterization} can be modified for continuous-alphabet channels to:
\begin{equation}
C(B)=\sup_{\substack{X:E[\rho_{\ell}(X)]\ge B,\\ \forall \ell\in\Theta}}\min_{\ell\in\Theta} I(X;Y(\ell)).
\end{equation}

Consider all $L$ channels having additive white Gaussian noise ${\cal N}(0,\sigma_\ell^2)$
and the energy function $b_{\ell}(y)=y^2$. 
Then the received energy at the $\ell$th receiver from input $x$ is
\begin{equation}
\rho_\ell(x)=\int_{-\infty}^{\infty}\frac{y^2}{\sigma_\ell\sqrt{2\pi}}\exp\left\{-\frac{(y-x)^2}{2\sigma_\ell^2}\right\}dy=x^2+\sigma_\ell^2,
\end{equation}
and the expectation is 
\begin{equation}
E[\rho_\ell(X)]=E[X^2+\sigma_\ell^2]=\rho_X^2+\sigma_\ell^2,
\end{equation}
where $\rho_X^2\triangleq E[X^2]=\int x^2dF(x)$.
We impose a peak power constraint so the input alphabet is 
$\cX=[-P,P]\subset\mathbb{R}$ and then denote the multicast capacity-energy function as $C(B, P)$.

Let $\cF_\ell$ be the space of all cdfs on the finite interval $[-P,P]$ such that, for any $F_\ell\in\cF_\ell$,
\begin{equation}
E_{F_\ell}[\rho_\ell(X)]\triangleq\int_{-P}^P x^2dF_\ell(x)+\sigma_\ell^2 \ge B.
\end{equation}
Since $\cF_\ell\subset\cF_{\ell'}$ if $\sigma_\ell^2\le\sigma_{\ell'}^2$, 
the multicast energy constraint \eqref{eqn:energy-constraint} is equivalent to
$
E[\rho_{\ell_{\min}}(X)]\ge B
$,
where $\ell_{\min}=\argmin_{\ell}\sigma_\ell^2$.
Also, since mutual information of a Gaussian channel decreases as the channel's variance increases, 
$\min_\ell I(X;Y(\ell))=I(X;Y(\ell_{\max}))$ where $\ell_{\max}=\argmax_{\ell}\sigma_\ell^2$.
Therefore, the multicast capacity-energy function can be rewritten as
\begin{equation}\label{eqn:energy-constraint-gaussian}
C(B,P)=\sup_{X\sim F\in\cF_{\ell_{\min}}}I(X;Y(\ell_{\max})).
\end{equation}

From \cite[Theorem 7]{Varshney2008}, 
the optimization problem \eqref{eqn:energy-constraint-gaussian} can be solved by the following theorem. 
\begin{thm}
There exists a unique capacity-energy achieving input $X_0\sim F_0$ 
and a constant $\lambda \ge 0$ such that
\begin{IEEEeqnarray}{rCl}
C(B,P)&=&\max_{X\sim F\in\cF_{\ell_{\min}}}I(X;Y(\ell_{\max}))-\lambda J(F)\\
&=&I(X_0;Y(\ell_{\max}))-\lambda J(F_0),
\end{IEEEeqnarray}
where 
\begin{equation}
J(F)\triangleq B-\sigma_{\ell_{\min}}^2-\int_{-P}^Px^2dF(x).
\end{equation}
Moreover, a necessary and sufficient condition for $F_0$ to achieve capacity-energy is 
\begin{multline}
\int_{-P}^P[i(x;F_0)+\lambda x^2]dF(x)\\\le I(X_0;Y(\ell_{\max}))+\lambda\int_{-P}^{P}x^2dF_0(x)
\end{multline}
for all $F\in\cF_{\ell_{\min}}$, where 
\begin{equation}
i(x;F)=\int Q_{\ell_{\max}}(y|x)\log\frac{Q_{\ell_{\max}}(y|x)}{p(y;F)}dy
\end{equation}
is the marginal information density \cite{Smith1971} and $p(y;F)$ is the output density function based on the input distribution $F$.
\end{thm}

\section{Segmentation Problem}
Let us consider the $L$-receivers segmentation problem in Fig.~\ref{fig:multicast_group}. 
Here, the receivers are partitioned into $K$ subsets, 
denoted as a segmentation $Q=\{\Theta(k)\}_{k=1}^K$ with $\bigcup_{k\in\{1,\cdots,K\}}\Theta(k)=\Theta$, 
and the same signal is transmitted to all receivers in the same group. 
As discussed in previous sections, the capacity of the $k$th group can be treated as 
a multicast system with $|\Theta(k)|$ receivers;
hence \eqref{eqn:capacity} applies to the multicast capacity for each group.

Given an energy constraint vector $\vec{B}$ for all $L$ receivers.
Let the set of all $\vec{B}$-admissible inputs for the $k$th group be $\cF_{\Theta(k)}=\bigcap_{\ell\in\Theta(k)}\cF_1(\ell)$,
then the corresponding capacity is 
\begin{equation}
C_{\Theta(k)}=\max_{X\in\cF_{\Theta(k)}}\min_{\ell\in\Theta(k)}I(X;Y(\ell)).
\end{equation}
A trivial performance characterization of a segmentation $Q$ is the worst capacity over all groups:
\begin{equation}\label{eqn:segmantationcapacity}
C_{Q}=\min_{k\in\{1,\ldots,K\}}C_{\Theta(k)}.
\end{equation}
Our objective is to find the segmentation $Q^*$ which maximizes \eqref{eqn:segmantationcapacity}, i.e.,
$Q^*=\argmax_{Q}C_{Q}$. 
This is a $k$-partition problem \cite{Chaimovich1993}, solvable by dynamic programming. 

A more interesting performance characterization is as follows.
Since the constraint of sending the same signal to the group reduces the capacity 
and the capacity is upper-bounded by removing this constraint (see discussion after Theorem \ref{thm:capacityfunction}),
the \emph{segmentation loss} for $k$th group can be defined as follows. 
\begin{equation}\label{eqn:segmentationloss}
\Delta_{\Theta(k)}=\left|C_{\Theta(k)}-\min_{m\in\Theta(k)}\max_{W\in\cF_{\Theta(k)}}I(W;Y(m))\right|.
\end{equation}
It is clear that $\Delta_{\Theta(k)}$ is zero when $|\Theta(k)|=1$. 
Then the objective is to find the segmentation $Q^*$ which minimizes the maximum $\Delta_{\Theta(k)}$ of all groups, i.e.,
\begin{equation}
Q^*=\argmin_{Q}\max_{k\in\{1,\ldots,K\}}\Delta_{\Theta(k)},
\end{equation}
which is also a minimax quantizer design problem \cite{VarshneyV2014c}. 

\begin{figure}
  \centering
    \includegraphics[width=3.5in]{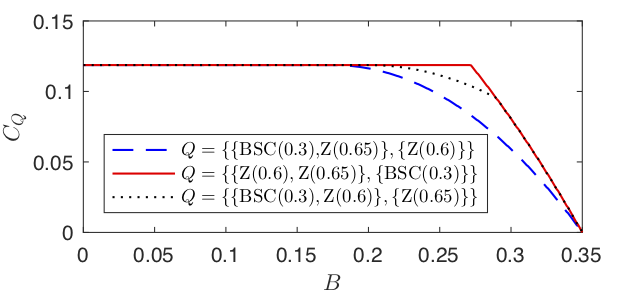}\vspace{-0.1in}
  \caption{An example of measurement \eqref{eqn:segmantationcapacity} for the
  three-receivers multicast communication ($L=3$) with two segmentations ($K=2$). 
  Three binary channels were considered: a BSC with crossover probability $0.3$ denoted by BSC($0.3$), 
  a Z-channel with a $1$ to $0$ probability $0.6$ denoted by Z($0.6$) and
  a Z-channel with a $1$ to $0$ probability $0.65$ denoted by Z($0.65$).}
  \vspace{-0.15in}
  \label{fig:ex-seg}
\end{figure}

\begin{figure}
  \centering
    \includegraphics[width=3.5in]{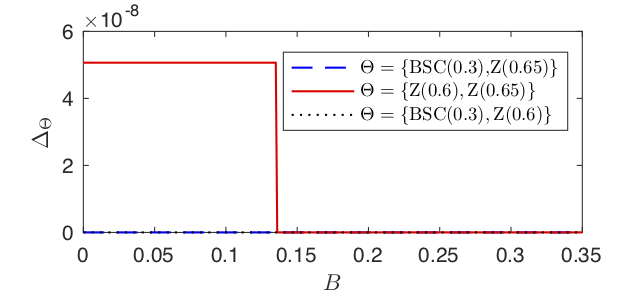}\vspace{-0.1in}
  \caption{An example of measurement \eqref{eqn:segmentationloss} for the
  same multicast communication in Fig.~\ref{fig:ex-seg}.
  Since $\Delta_\Theta=0$ when $|\Theta|=1$, we only consider the group of size $2$.}
  \vspace{-0.15in}
  \label{fig:ex-segloss}
\end{figure}

An example of clustering three receivers into two groups is given in Figs.~\ref{fig:ex-seg} and \ref{fig:ex-segloss}, 
in which a BSC($\epsilon=0.3$) and two Z-channels with $\epsilon_0=\{0.6,0.65\}$ were considered. 
Fig.~\ref{fig:ex-seg} shows \eqref{eqn:segmantationcapacity} for every segmentation,
indicating that grouping two Z-channels yields the optimal capacity. 
However Fig.~\ref{fig:ex-segloss} shows the opposite, when using segmentation loss \eqref{eqn:segmentationloss}, 
grouping the BSC with any other Z-channel yields zero segmentation loss.
Hence, in this example, the grouping with the optimal capacity is not best when segmentation loss is the main concern. 

\section{conclusion}
This paper discussed the simultaneous multicast of energy and information, where the effective channel is equivalent to a constrained compound channel. We studied the capacity for this channel, and derived a single-letter capacity expression under a received energy constraint.
The segmentation problem was also discussed to analyze the situation where receivers are partitioned into related groups. An interesting avenue for future work is to extend the code constructions in \cite{TandonMV2016} and \cite{WuTVM2017} to provide explicit constrained codes for multicast SIET.   

\appendices
\section{Unequal Energy Constraints}
\label{append:letterization_general}
To show that the single-letter form can be also derived for unequal energy constraints 
$\vec{B}=(B_1,\ldots,B_L)$ and harvest functions $\vec{b}=(b_1,\ldots,b_L)$,   
we first prove the domain ${\cal B}$ in \eqref{def:domain} is convex. 
\begin{thm}
If $\vec{B}_1=((B_1)_1,(B_1)_2, \ldots, (B_1)_L)$ and $\vec{B}_2=((B_2)_1,(B_2)_2, \ldots, (B_2)_L)$ are both in ${\cal B}$, 
then $(\alpha\vec{B}_1+\beta\vec{B}_2)\in{\cal B}$ for all $\alpha\in[0,1]$ and $\beta=1-\alpha$. 
\end{thm}
\begin{IEEEproof}
Let $\vec{q}_1=\left((q_1)_1,(q_1)_2,\ldots,(q_1)_{|{\cal X}|}\right)$ and $\vec{q}_2=\left((q_2)_1,(q_2)_2,\ldots,(q_2)_{|{\cal X}|}\right)$ be nonnegative row vectors such that 
$\sum_{i=1}^{|{\cal X}|}(q_{j})_i=1$ and
$\vec{q}_j\cdot[P(\ell)]\cdot\vec{b}_\ell=(B_j)_\ell$ for all $j\in\{1,2\}$ and $\ell\in\Theta$.
We have 
$(\alpha\vec{q}_1+\beta\vec{q}_2)\cdot[P(\ell)]\cdot\vec{b}_\ell=(B_1)_\ell+(B_2)_\ell$ for all $\ell\in\Theta$.
Since $\sum_{i=1}^{|{\cal X}|}(\alpha(q_1)_i+\beta(q_2)_i)=\alpha+\beta=1$, 
we conclude that ${\cal B}$ is convex. 
\end{IEEEproof}

The following theorem that the $n$th multicast capacity-energy function for unequal constraints can be evaluated as a single-letter expression in a similar way. 
\begin{thm}\label{thm:letterization}
For any DMC, $C_n(\vec{B})=nC_1(\vec{B})$ for all $n\in\mathbb{Z}^+$ and $\vec{B}\in{\cal B}$.
\end{thm}
\begin{IEEEproof}
To prove the equality, we start with the case of ``$\le$''.
Let $X_1^n$ be the channel input with the corresponding outputs $\{Y(\ell)_1^n\}_{\ell\in\Theta}$ which achieves $C_n(\vec{B})$, 
then we have $E[b_\ell(Y(\ell)_1^n)]\ge nB_\ell$ for all $\ell\in\Theta$. 
We also let $X^*$ be the channel input with the corresponding outputs $\{Y^*(\ell)\}_{\ell\in\Theta}$ such that 
$p_{X^*}=\sum_{i=1}^n \frac{1}{n}p_{X_i}$.
Due to the memoryless, We have 
\begin{IEEEeqnarray}{rCl}
E[b_\ell(Y^*(\ell))]&=&\frac{1}{n}\sum_{i=1}^nE[b_\ell(Y(\ell)_i)]\\
&=&\frac{1}{n}E[b_\ell(Y(\ell)_1^n)]\ge B_\ell
\end{IEEEeqnarray}
and $I(X_1^n;Y(\ell)_1^n)\le\sum_{i=1}^nI(X_i;Y(\ell)_i)$ for all $\ell\in\Theta$.
Hence
\begin{IEEEeqnarray}{rCl}
C_n(\vec{B})&=&\min_{\ell\in\Theta}I(X_1^n;Y(\ell)_1^n)\\
&\le&\min_\ell \sum_{i=1}^nI(X_i;Y(\ell)_i)\\
&=&n\min_\ell \frac{\sum_{i=1}^nI(X_i;Y(\ell)_i)}{n}\\
&\le&n\min_\ell I(X^*;Y^*(\ell))\\
&\le&nC_1(\vec{B}).
\end{IEEEeqnarray}

For the case of ``$\ge$'',
we let $X$ be the channel input with the distribution $p_X$ and the corresponding outputs $\{Y(\ell)\}_{\ell\in\Theta}$ achieving $C_1(\vec{B})$.
A sequence of channel inputs $X_1^n=(X_1,X_2,\cdots,X_n)$ is created, in which each $X_i$ is i.i.d. to $X$. 
Let $\{Y(\ell)_i^n\}_{\ell\in\Theta}$ be the corresponding channel outputs of $X_i^n$.
By definition, $E[b_\ell(Y(\ell)_1^n)]=nE[b_\ell(Y(\ell))]\ge nB_\ell$ for all $\ell\in\Theta$.
Therefore
\begin{IEEEeqnarray}{rCl}
C_n(\vec{B})&\ge&\min_{\ell\in\Theta}I(X_1^n;Y(\ell)_1^n)\\
&=&\min_{\ell\in\Theta}\sum_{i=1}^nI(X;Y(\ell))\\
&=&nC_1(\vec{B}).
\end{IEEEeqnarray}
\end{IEEEproof}

\bibliographystyle{IEEEtran} 
\bibliography{abrv,conf_abrv,lrv_lib}

\newcommand{\SortNoop}[1]{}
\begin{thebibliography}{10}
\providecommand{\url}[1]{#1}
\csname url@samestyle\endcsname
\providecommand{\newblock}{\relax}
\providecommand{\bibinfo}[2]{#2}
\providecommand{\BIBentrySTDinterwordspacing}{\spaceskip=0pt\relax}
\providecommand{\BIBentryALTinterwordstretchfactor}{4}
\providecommand{\BIBentryALTinterwordspacing}{\spaceskip=\fontdimen2\font plus
\BIBentryALTinterwordstretchfactor\fontdimen3\font minus
  \fontdimen4\font\relax}
\providecommand{\BIBforeignlanguage}[2]{{%
\expandafter\ifx\csname l@#1\endcsname\relax
\typeout{** WARNING: IEEEtran.bst: No hyphenation pattern has been}%
\typeout{** loaded for the language `#1'. Using the pattern for}%
\typeout{** the default language instead.}%
\else
\language=\csname l@#1\endcsname
\fi
#2}}
\providecommand{\BIBdecl}{\relax}
\BIBdecl

\bibitem{TongLWZ2010}
B.~Tong, Z.~Li, G.~Wang, and W.~Zhang, ``How wireless power charging technology
  affects sensor network deployment and routing,'' in \emph{Proc. 2010 IEEE
  Int. Conf. Distrib. Comput. Syst. (ICDCS)}, Jun. 2010, pp. 438--447.

\bibitem{LeeLL2012}
D.-S. Lee, Y.-H. Liu, and C.-R. Lin, ``A wireless sensor enabled by wireless
  power,'' \emph{Sensors}, vol.~12, no.~12, pp. 16\,116--16\,143, Nov. 2012.

\bibitem{ZhangMH2015}
R.~Zhang, R.~G. Maunder, and L.~Hanzo, ``Wireless information and power
  transfer: From scientific hypothesis to engineering practice,'' \emph{{IEEE}
  Commun. Mag.}, vol.~53, no.~8, pp. 99--105, Aug. 2015.

\bibitem{ClerckxZSNKP2018}
B.~Clerckx, R.~Zhang, R.~Schober, D.~W.~K. Ng, D.~I. Kim, and H.~V. Poor,
  ``Fundamentals of wireless information and power transfer: From rf energy
  harvester models to signal and system designs,'' arXiv:1803.07123 [cs.IT].,
  Mar. 2018.

\bibitem{Varshney2008}
L.~R. Varshney, ``Transporting information and energy simultaneously,'' in
  \emph{Proc. 2008 IEEE Int. Symp. Inf. Theory}, Jul. 2008, pp. 1612--1616.

\bibitem{GroverS2010}
P.~Grover and A.~Sahai, ``{S}hannon meets {T}esla: Wireless information and
  power transfer,'' in \emph{Proc. 2010 IEEE Int. Symp. Inf. Theory}, Jun.
  2010, pp. 2363--2367.

\bibitem{Varshney2012}
L.~R. Varshney, ``On energy/information cross-layer architectures,'' in
  \emph{Proc. 2012 IEEE Int. Symp. Inf. Theory}, Jul. 2012, pp. 1361--1365.

\bibitem{LiuZC2013}
L.~Liu, R.~Zhang, and K.-C. Chua, ``Wireless information transfer with
  opportunistic energy harvesting,'' \emph{{IEEE} Trans. Wireless Commun.},
  vol.~12, no.~1, pp. 288--300, Jan. 2013.

\bibitem{TandonMV2016}
A.~Tandon, M.~Motani, and L.~R. Varshney, ``Subblock-constrained codes for
  real-time simultaneous energy and information transfer,'' \emph{{IEEE} Trans.
  Inf. Theory}, vol.~62, no.~7, pp. 4212--4227, Jul. 2016.

\bibitem{WuTVM2017}
T.-Y. Wu, A.~Tandon, L.~R. Varshney, and M.~Motani, ``Skip-sliding window
  codes,'' arXiv:1711.09494 [cs.IT]., Nov. 2017.

\bibitem{FouladgarS2012}
A.~M. Fouladgar and O.~Simeone, ``On the transfer of information and energy in
  multi-user systems,'' \emph{{IEEE} Commun. Lett.}, vol.~16, no.~11, pp.
  1733--1736, Nov. 2012.

\bibitem{ZhangH2013}
R.~Zhang and C.~K. Ho, ``{MIMO} broadcasting for simultaneous wireless
  information and power transfer,'' \emph{{IEEE} Trans. Wireless Commun.},
  vol.~12, no.~5, pp. 1989--2001, May 2013.

\bibitem{TallaKRNSG2017}
V.~Talla, B.~Kellogg, B.~Ransford, S.~Naderiparizi, J.~R. Smith, and
  S.~Gollakota, ``Powering the next billion devices with {Wi-Fi},''
  \emph{Commun. ACM}, vol.~60, no.~3, pp. 83--91, Mar. 2017.

\bibitem{LapidothN1998}
A.~Lapidoth and P.~Narayan, ``Reliable communication under channel
  uncertainty,'' \emph{{IEEE} Trans. Inf. Theory}, vol.~44, no.~6, pp.
  2148--2177, Oct. 1998.

\bibitem{ZengCZ2017}
Y.~Zeng, B.~Clerckx, and R.~Zhang, ``Communications and signals design for
  wireless power transmission,'' \emph{{IEEE} Trans. Commun.}, vol.~65, no.~5,
  pp. 2264--2290, May 2017.

\bibitem{ElGamalK2011}
A.~El~Gamal and Y.-H. Kim, \emph{Network Information Theory}.\hskip 1em plus
  0.5em minus 0.4em\relax Cambridge: Cambridge University Press, 2011.

\bibitem{WuTVM2018_arXiv}
T.-Y. Wu, A.~Tandon, L.~R. Varshney, and M.~Motani, ``Multicasting energy and
  information simultaneously,'' arXiv:1806.11271 [cs.IT]., Jun. 2018.

\bibitem{Smith1971}
J.~G. Smith, ``The information capacity of amplitude- and variance-constrained
  scalar {G}aussian channels,'' \emph{Inf. Control}, vol.~18, no.~3, pp.
  203--219, Apr. 1971.

\bibitem{Chaimovich1993}
M.~Chaimovich, ``Fast exact and approximate algorithms for k-partition and
  scheduling independent tasks,'' \emph{Discrete Math.}, vol. 114, no. 1--3,
  pp. 87--103, Apr. 1993.

\bibitem{VarshneyV2014c}
K.~R. Varshney and L.~R. Varshney, ``Optimal grouping for group minimax
  hypothesis testing,'' \emph{{IEEE} Trans. Inf. Theory}, vol.~60, no.~10, pp.
  6511--6521, Oct. 2014.

\end{thebibliography}

\end{document}